\documentclass[11pt]{article}
\usepackage{graphicx}
\usepackage{amsmath}
\usepackage{amssymb}

\begin{document}

\title{de Sitter geodesics in stereographic charts}

\author{{Ion I. Cot\u aescu\footnote{E-mail: i.cotaescu@e-uvt.ro}}\\
{\small \it West University of Timi\c{s}oara,}\\
{\small \it V.  P\^{a}rvan Ave.  4, RO-300223 Timi\c{s}oara, Romania}}

\maketitle

\begin{abstract}
It is shown that the stereographic coordinates on the  $(1+3)$-dimensional de Sitter spacetime induce, in the parametric equations of all the timelike geodesics, artificial singularities which do not have any physical motivation. 
\end{abstract}


\newpage

\section{Introduction}

The free geodesic motion on the  $(1+3)$-dimensional de Sitter (dS) spacetime was considered by may authors which studied the conserved quantities  \cite{cac1,cac2} and their role in determining the geodesic trajectories \cite{CdS1,CdS2,Cnew}. 

Another attractive direction is the so called dS relativity developed in dS charts with stereographic coordinates where the $SO(1,3)$ subgroup of the dS isometry group $SO(1,4)$ generates linear isometries as in special relativity \cite{A1,A2,P}.

In this short paper we would like to show that the stereographic charts are not suitable for  physical interpretation because of a specific mathematical behavior which induces artificial singularities in the parametric equations of the timelike geodesics. 

We start in the second section presenting the timelike geodesics in terms of conserved quantities in the conformal charts of the dS manifold. In the next section we derive the parametric equations of these geodesics in stereographic coordinates pointing out the singularties which do not have a physical meaning.  The final conclusion is that the stereographic charts are not suitable for doing physics.

\section{Geodesics in conformal charts}

The de Sitter spacetime $(M,g)$ is defined as a hyperboloid of radius $1/\omega$  in the five-dimensional flat spacetime $(M^5,\eta^5)$ of coordinates $z^A$  (labeled by the indices $A,\,B,...= 0,1,2,3,4$) having the pseudo-Euclidean metric $\eta^5={\rm diag}(1,-1,-1,-1,-1)$. The local charts $\{x\}$  of coordinates $x^{\mu}$ ($\alpha,\mu,\nu,...=0,1,2,3$) can be introduced on $(M,g)$ giving the set of functions $z^A(x)$ which solve the hyperboloid equation,
\begin{equation}\label{hip}
\eta^5_{AB}z^A(x) z^B(x)=-\frac{1}{\omega^2}\,,
\end{equation}
where  $\omega$ denotes the Hubble de Sitter constant since in our notations  $H$ is reserved for the energy (or Hamiltonian) operator \cite{CGRG}. 

In what follows we use only Cartesian space coordinates which satisfy the condition  $z^i\propto x^i$ such that at least the $SO(3)$ symmetry becomes global, any quantity bearing space indices transforming under rotations as $SO(3)$ vectors or tensors. Under such circumstances, we may use the vector notation for the $SO(3)$ vectors, including the position one, ${\bf x}=(x^1,x^2,x^3)\in {\mathbb R}^3$. The norms of the conserved vectors have to be denoted simply as $V=|{\bf V}|$.  

We start  with the {conformal} charts, $\{t_c,{\bf x}_c\}$, with the conformal time $t_c$ and Cartesian spaces coordinates $x^i_c$ ($i,j,k,...=1,2,3$), defined by the functions 
\begin{eqnarray}
z^0(x_c)&=&-\frac{1}{2\omega^2 t}\left[1-\omega^2({t}_c^2 - {\bf x}_c^2)\right]\,,
\nonumber\\
z^i(x_c)&=&-\frac{1}{\omega t}x^i_c \,, \label{Zx}\\
z^4(x_c)&=&-\frac{1}{2\omega^2 t}\left[1+\omega^2({t}_c^2 - {\bf x}_c^2)\right]\,.
\nonumber
\end{eqnarray}
These charts  cover the expanding part of $M$ for $t_c \in (-\infty,0)$
and ${\bf x}_c\in {\mathbb R}^3$ while the collapsing part is covered by
similar charts with $t_c >0$. We stress that here we restrict ourselves to consider only the expanding portion with $t_c<0$ related to the proper time $t_p$ as   \cite{BD}
\begin{equation}
t_c=-\frac{1}{\omega}\,e^{-\omega t_p}\,, \quad t_p\in {\Bbb R}
\end{equation}
such that  the origin $t_p=0$ correspond to $t_c=-\frac{1}{\omega}$. 

We have shown that in this chart the timelike geodesic of a particle of mass $m$ and momentum ${\bf P}$ has the rectilinear trajectory  \cite{CGRG},
\begin{equation}\label{geo}
{\bf x}_c(t)={\bf x}_{c0}+\frac{{\bf P}}{\omega {P}^
2} \left(\sqrt{m^2+{P}^{2}\omega^2 t_{c0}^2}-\sqrt{ m^2+{P}^2
\omega^2 t_c^2}\, \right)\,,
\end{equation}
which is completely determined by the initial condition  ${\bf x}_c(t_{c0}) ={\bf x}_{c0}$  and the conserved momentum ${\bf P}$.  
The geodesc is in a plane, orthogonal on the angular momentum ${\bf  L}$, where it is convenient to consider  the orthonormal basis $\{{\bf n}_{\perp},{\bf n}_P\}$ formed by  ${\bf n}_P=\frac{{\bf P}}{P}$ and its orthogonal complement, ${\bf n}_{\perp}$. In this frame we use the local Cartesian coordinates $(x_{\perp}, x_{\parallel})$ such that any position vector can be written as ${\bf x}={\bf x}_{\perp}+{\bf x}_{\parallel}=x_{\perp} {\bf n}_{\perp}+x_{\parallel} {\bf n}_{P}$.   The advantage of this basis is that now we can express the geodesic equation in terms of conserved quantities as \cite{Cnew}
\begin{eqnarray}
{\bf x}_c(t_c)&=&{\bf x}_{cA}+{\bf n}_P\,{x}_{c\parallel}(t_c) \nonumber\\
& =&{\bf n}_{\perp}\frac{L}{P}+{\bf n}_P\frac{1}{\omega P}\left(E-\sqrt{m^2+P^2\omega^2 t_c^2}\right)\,,\label{geocon}
\end{eqnarray}
where $E$ is the energy  while ${\bf x}_{cA}$ gives the position of the closest point to origin \cite{Cnew}. The function ${x}_{c\parallel}(t_c)$ describes the motion along the direction ${\bf n}_P$ between the limits
\begin{equation}\label{dom1}
-\infty<{x}_{c\parallel}(t_c)\le \frac{E-m}{\omega P}\,,\quad E\ge m\,,
\end{equation}
since $-\infty<t_c\le 0$.

\section{Geodesics in stereographic charts}

An interesting type of local coordinates $\{t,\vec{x}\}$ may be introduced by the embedding functions 
\begin{eqnarray}
z^{\mu}(x)&=&\Omega(s) x^{\mu}\,,\\
z^4(x)&=&-\frac{1}{\omega}\sqrt{1+\omega^2 s^2 \Omega(s)^2}\,,
\end{eqnarray}
where $s=\sqrt{t^2-{{\bf x}\,}^2}$. In this case the Lorentz symmetry becomes global since these coordinates transform under Lorentz isometries according to the linear transformations as in special relativity. This property inspired the so called de Sitter relativity 
which was built in the stereographic chart where the function $\Omega$ has the special form
\begin{equation}
\Omega(x)=\frac{1}{1-\frac{1}{4}\omega^2 s^2}
\end{equation} 
producing the conformal flat line element with $g=\Omega(x)^2{\rm diag}(1,-1,-1,-1)$  \cite{A1,A2}.

The stereographic coordinates can be related to the conformal ones by solving the system 
$z(x_c)=z(x)$ that yields,
\begin{eqnarray}
t(t_c,{\bf x}_c)&=&\frac{2}{\omega}\,\frac{\omega^2(t_c^2-{{\bf x}_c}^2)-1}{\omega^2(t_c^2-{{\bf x}_c}^2)+2\omega t_c+1}\,,\label{tst}\\
x^i(t_c,{\bf x}_c)&=&-\frac{2}{\omega}\,\frac{4x^i_c}{\omega^2(t_c^2-{{\bf x}_c}^2)+2\omega t_c+1}\,,\label{xst}
\end{eqnarray}
such that
\begin{equation}
s^2=\frac{4}{\omega^2}\,\frac{\omega^2(t_c^2-{{\bf x}_c}^2)-2\omega t_c+1}{\omega^2(t_c^2-{{\bf x}_c}^2)+2\omega t_c+1}\,.
\end{equation}
We observe that all these functions are singular for $|{\bf x}_c|=|t_c+\frac{1}{\omega}|$ which is just the condition of defining the null cone at $t_c=-\frac{1}{\omega}$ of an observer staying at rest in the origin of the conformal chart, having the worldline along the vector field $\partial_{t_c}$.  Moreover, the events of this worldline, $(t_c,0)$, are mapped into those of the corresponding worldline in the stereographic chart, $(t_*,0)$, where the time
\begin{equation}
t_*=\frac{2}{\omega}\,\frac{\omega t_c-1}{\omega t_c+1}=-\frac{2}{\omega}\coth \frac{\omega t_p}{2}
\end{equation}
is singular just for $t_c=-\frac{1}{\omega}$ which corresponds to the origin of the proper time $t_p=0$. Thus, the mapping from the conformal chart to the stereographic one is singular on the null cone of the conformal chart mapping the present of the conformal chart into the sterographic time $-\infty$.

This unwanted behavior may affect the geodesic equations in stereograpphic charts which also can be singular.  In order to convince that,  let us consider the timelike geodesic (\ref{geocon}) in the plane  $({\bf n}_{\perp},{\bf n}_P)$ mapped,  according to Eqs. (\ref{tst}) and (\ref{xst}), into a trajectory in the same plane having the parametric equations
\begin{eqnarray}
t(\tau)&=&\frac{2}{\omega}\,\frac{\omega^2L^2+E^2+m^2+P^2-2E\sqrt{m^2+P^2\tau^2}}{\omega^2 L^2+E^2+m^2- P^2(2\tau+1)-2E\sqrt{m^2+P^2\tau^2}}\,,\\
x_{\perp}(\tau)&=&\frac{4LP}{\omega^2 L^2+E^2+m^2- P^2(2\tau+1)-2E\sqrt{m^2+P^2\tau^2}}\,,\\
x_{\parallel}(\tau)&=&\frac{4P}{\omega}\,\frac{E-\sqrt{m^2+P^2\tau^2}}{\omega^2 L^2+E^2+m^2- P^2(2\tau+1)-2E\sqrt{m^2+P^2\tau^2}}\,,
\end{eqnarray}
where $\tau=\omega t_c \in (-\infty,0]$. As expected, all these equations are singular as we can see from the numerical examples presented in Fig. 1.  

{ \begin{figure}
  \centering
    \includegraphics[scale=0.38]{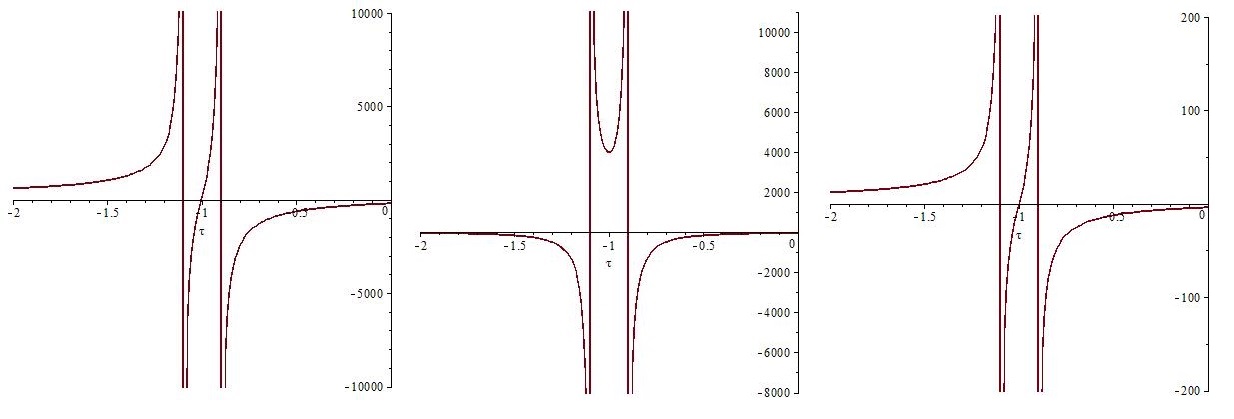}
    \caption{The functions $t(\tau)$, $x_{\perp}(\tau)$  and $x_{\parallel}(\tau)$  on the domain  $-2\le\tau \le 0$ for $\omega=0.01$, $m=0.05$, $P=0.001$ and $L=10 P$.}
  \end{figure}}

\section{Conclusion}

We have seen that all the geodesic equations in the stereographic chart are singular while in other charts these are continuous or even analytic. This means that these singularities are induced mathematically by the special choice of the stereographic coordinates without reflecting a physical reality.  For this reason, the conclusion is that the stereographic charts are not suitable for physical interpretations.

\subsection*{Acknowledgements}

This work was partially supported by a grant of the Ministry of National Education and Scientific Research, RDI Programme for Space Technology and Advanced Research - STAR, project number 181/20.07.2017.

\end{document}